\documentclass[10pt, journal, twocolumn, oneside, final, romanappendices]{IEEEtran}

\usepackage{color}
\usepackage{multirow}
\usepackage{graphicx}
\usepackage{epstopdf}
\usepackage[cmex10]{amsmath}
\usepackage{amssymb}

\usepackage{multirow}
\usepackage{amsthm}
\usepackage{cite}
\usepackage{acronym} 
\usepackage{algorithm, algcompatible}

\algnewcommand\INPUT{\item[\textbf{Input:}]}
\algnewcommand\OUTPUT{\item[\textbf{Output:}]}

\allowdisplaybreaks

\makeatletter
\newcommand{\doublewidetilde}[1]{{%
  \mathpalette\double@widetilde{#1}%
}}
\newcommand{\double@widetilde}[2]{%
  \sbox\z@{$\m@th#1\widetilde{#2}$}%
  \ht\z@=.9\ht\z@
  \widetilde{\box\z@}%
}
\makeatother

\begin{document}

\title{A New Interweave Cognitive Radio Scheme for Out-band Energy Harvesting Systems}

\author{Nikolaos~I.~Miridakis and Theodoros~A.~Tsiftsis,~\IEEEmembership{Senior Member,~IEEE}
\thanks{N. I. Miridakis is with the Department of Computer Systems Engineering, Piraeus University of Applied Sciences, 12244, Aegaleo, Greece (e-mail: nikozm@unipi.gr).}
\thanks{T. A. Tsiftsis is with the School of Engineering, Nazarbayev University, Astana 010000, Kazakhstan (e-mail: theodoros.tsiftsis@nu.edu.kz).}
}


\maketitle

\begin{abstract}
In this paper, a new interweave cognitive radio (CR) transmission scheme is analytically presented for energy harvesting (EH)-enabled transmitting nodes. EH is performed via a dedicated power beacon (PB) base station through wireless power transfer, which harvests energy to both primary and secondary systems. The out-band mode-of-operation between PB and the two heterogeneous systems is considered (i.e., data communication and EH are realized at different frequency bands). The performance of the new interweave CR system is studied under independent and not necessarily identically distributed $\kappa-\mu$ shadowed faded channels. Representative numerical and simulation results corroborate the effectiveness of the proposed scheme and presented analysis, while some useful engineering insights are provided.
\end{abstract}

\begin{IEEEkeywords}
Cognitive radio (CR), energy harvesting (EH), interweave secondary transmission, performance analysis.
\end{IEEEkeywords}

\IEEEpeerreviewmaketitle

\section{Introduction}
\IEEEPARstart{E}{nergy} harvesting (EH) via wireless power transfer can be used for enabling perpetual operation of wireless devices. Recently, EH has attracted increasing interests, mainly due to its capability of converting received RF signals into electricity, which in turn is able to provide stable and controllable power to prolong the lifetime of low-power energy-constrained autonomous networks \cite{j:KansalSadaf2007,j:FalkensteinCostinett2011,j:XiaAissa2015}. Typical examples are machine-type communications such as the Internet of Things, Internet of Everything, Smart X, etc., where \emph{connectivity rather than high throughput} seems to be of prime importance. On another front, cognitive radio (CR) has emerged as one of the most promising technologies to resolve the issue of spectrum scarcity. The key enabling technology of CR relies on the provision of capability to share the spectrum in an opportunistic manner, such that the secondary transmissions do not cause any harmful interference to the primary communication \cite{j:Haykin}.

The volume of the aforementioned machine-type communication devices (e.g., sensors) is expected to be massive and widespread worldwide to support various upcoming $5$G-and-beyond applications \cite{j:BockelmannPratas2016}. Thereby, spectrum- and energy-efficiency should be some of the most prominent building blocks for the design of such systems. Consequently, CR and EH can be jointly utilized to achieve this goal due to their complementary benefits \cite{j:HuangHan2015,j:MiridakisTsiftsisAlexandropoulosDebbah2016}.

Essentially, systems in which EH and data transfer occur at different frequency bands are known as out-band systems, while those working at the same frequencies are known as in-band systems. Obviously, out-band systems are interference-free between the power and data transfer segments \cite{j:XiaAissa2015}. To date, most related research works are focused either on the underlay CR transmission mode (i.e., CR and primary communication overlaps to one another using the same resources) and/or the in-band EH operation (i.e., see \cite{j:KuLiChen2016} and relevant references therein).  

In this paper, we propose a new interweave CR system, which operates under the presence of a primary one. We consider the case when both systems use EH-enabled transmission via a certain power beacon (PB) base station, which transmits in a dedicated frequency band (i.e., an out-band EH transmission). Notably, the out-band transmission mode is less complex than its corresponding in-band counterpart \cite{j:LeeZhang2017}. The main benefits of the proposed scheme are: (\emph{i}) no co-channel interference is produced between the two systems; (\emph{ii}) the CR system is purely EH-enabled since only an EH circuit and a micro-supercapacitor (as an energy storage device) are required, which is a much more cost-effective solution than rechargeable batteries \cite{j:MorsiMichalopoulosSchober2018}. Moreover, some useful performance metrics are derived in closed-form when the signals undergo $\kappa-\mu$ shadowed fading so as to approach realistic channel conditions \cite{j:MartinezParisJerez2017,j:PlazaolaParis2017} along with some impactful engineering insights. 

{\bf Notation}: $|x|$ takes the absolute value of $x$ while $\|\mathbf{x}\|$ is the Euclidean norm of vector $\mathbf{x}$. $\mathbb{E}[\cdot]$ is the expectation operator and ${\rm Pr}[\cdot]$ returns probability. The symbol $\overset{\text{d}}=$ means equality in distribution. The functions $f_{X}(\cdot)$, $F_{X}(\cdot)$ and $\overline{F}_{X}(\cdot)$ represent probability density function (PDF), cumulative distribution function (CDF) and complementary CDF of the random variable (RV) $X$, respectively. Complex-valued Gaussian RVs with mean $\mu$ and variance $\sigma^{2}$ are denoted as $\mathcal{CN}(\mu,\sigma^{2})$. Finally, $\Gamma(\cdot)$ denotes the gamma function \cite[Eq. (8.310.1)]{tables}, $\Gamma(\cdot,\cdot)$ is the upper incomplete gamma function \cite[Eq. (8.350.2)]{tables}, and $\psi(\cdot)$ is the digamma function \cite[Eq. (8.360.1)]{tables}.

\section{System and Channel Model}
\label{System Model}
\begin{figure}[!t]
\centering
\includegraphics[keepaspectratio,width=3.0in]{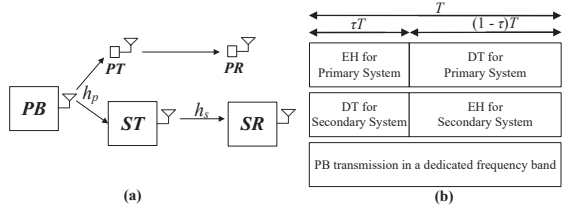}
\caption{(a) The considered system configuration. The parameters $h_{\{p,s\}}$ denote the involved channel gains, which are explicitly defined hereinafter; (b) The mode-of-operation for the proposed EH interweave CR scheme.}
\label{fig1}
\end{figure}

Consider a point-to-point (secondary) CR system, which operates under the presence of a primary system, as illustrated in Fig.~\ref{fig1}a, where PT and PR denote the single-antenna primary Tx and Rx, respectively. Similarly, ST and SR denote the single-antenna secondary Tx and Rx, respectively. It is assumed that the Tx nodes of both systems harvest energy by a dedicated PB base station, which is in their close vicinity. Afterwards, based on this energy, they can transmit their data streams to their corresponding Rx nodes. Independent and not necessarily identically distributed $\kappa-\mu$ shadowed faded channels are assumed for the various involved links to cope with realistic conditions. The reason why the $\kappa-\mu$ shadowed distribution is adopted for the channel modeling is threefold; generality, tractability and accuracy. In fact, this model includes other popular distributions as special cases, such as Rician, Nakagami-$m$ and Rayleigh fading models. Most importantly, it provides an improved fit to field measurements than the latter models and a relative mathematical simplicity at the same time \cite{j:PlazaolaParis2017}. 

The communication activity of the single-antenna primary system occurs in consecutive time frames, each with a fixed duration $T$. The primary Tx uses a certain fraction of the frame with duration $\tau T$ for EH and then utilizes the remaining time $(1-\tau)T$ for data transmission (in a similar basis as in, e.g., \cite{j:ZhangChen2017,j:ZhaoLinKaragiannidis2018}). In out-band EH systems, PB operates in a dedicated frequency band (different from the one used for data communication).\footnote{As an illustrative example, typical frequencies that are used for prototype development of far-field wireless power transfer systems are $2.45$GHz and $5.8$GHz \cite{j:FalkensteinCostinett2011}.} Capitalizing on this approach, the secondary system operates in an interweave CR basis, such that it causes no interference onto the primary system. To this end, the secondary system uses the former time fraction for data communication and the latter one for EH (see Fig.~\ref{fig1}b). In what follows, we focus on the analysis and evaluation of the secondary system performance, which is the scope of current work, given a fixed switching time $\tau$ specified by the primary system.\footnote{The switchning time $\tau$ can be computed by the primary system without taking into account any underlying secondary activity, e.g., as in \cite{j:ZhaoLinKaragiannidis2018}. Afterwards, $\tau$ is captured by the secondary system via feedback signaling, which can essentially be used to inform the involved primary nodes.} 

During the EH phase of the $n^{\rm th}$ time instance, the received signal at ST, $y_{p}$, reads as
\begin{align}
y_{p}[n]=\sqrt{P_{b}d^{-\alpha}_{\rm PBST}}h_{p}[n]s_{\rm PB}+n_{\rm ST}[n],
\label{signalST}
\end{align}
where $d_{\rm PBST}$ and $\alpha$ is the distance and path-loss factor of the $\rm PB-ST$ channel link, respectively. Also, $h_{p}[n]$, $s_{\rm PB}$ and $n_{\rm ST}[n]$ denote, respectively, the channel fading, reference signal and additive white Gaussian noise (AWGN) at ST, regarding the $\rm PB-ST$ channel link. 

In current work, a `\emph{one-shot}' policy is adopted; the data transmission phase occurs in a given time instance only when the energy buffer has been full ($=M T$ Joules) in the previous time instance(s), where $M$ is the predetermined Tx power at ST, while $T M$ denotes the total capacity of the energy storage circuit at ST. Then, it uses the required energy to transmit during $\tau T$ fraction of time (i.e., using a predetermined Tx power $M$ to meet the appropriate transmission quality level) in a single shot. Thus, the corresponding harvested energy at the $n^{\rm th}$ time instance is given by
\begin{align}
{\rm EH}[n]\triangleq \min\left\{\frac{\eta (1-\tau)T P_{b}|h_{p}|^{2}[n]}{d^{\alpha}_{\rm PBST}},\tau T M\right\},
\label{EH}
\end{align}
where $0<\eta \leq 1$ represents the power-to-energy conversion efficiency factor. Studying a finite-size energy buffer, supercapacitors can be leveraged since they provide quite a fast charging rate, long cycle lifetime, high storage efficiency, while they can be integrated on chip, which makes them suitable storage devices for EH nodes. \cite{j:KestutisKeskinen2016,j:KansalSadaf2007}. 

Afterwards, based on a full energy buffer at the $n^{\rm th}$ time instance, the secondary system enters the data transmission phase in the $(n+1)^{\rm th}$ time instance. Hence, the received signal at SR, conditioned on the fact that ${\rm EH}[n]=M T$, is expressed as
\begin{align}
y_{s}[n+1]=\frac{\sqrt{M}}{d^{\alpha_{s}}_{\rm STSR}} h_{s}[n+1]s_{\rm s}[n+1]+n_{\rm SR}[n+1],
\label{signalSR}
\end{align}
where $d_{\rm STSR}$ and $\alpha_{s}$ is the distance and path-loss factor of the $\rm ST-SR$ channel link, respectively. Also, $h_{s}[n+1]$, $s_{\rm s}[n+1]$ and $n_{\rm SR}[n+1]$ denote, respectively, the channel fading, unit-energy transmitted symbol and AWGN at SR, regarding the $\rm ST-SR$ channel link. Otherwise, it transmits no data during $\tau T[n+1]$, whereas it enters to the EH phase for (at least) a duration of $T[n+1]$ instead.\footnote{Hereinafter, for notational simplicity, we use a normalized time duration by setting $T=1$.} 

Moreover, it is assumed that $\{n_{\rm ST},n_{\rm SR}\}\overset{\text{d}}=\mathcal{CN}(0,N_{0})$, as time-invariant random processes. Also, the channel fading gains are modeled as $\kappa-\mu$ shadowed RVs, such that
\begin{align}
|h_{v}|^{2}\overset{\text{d}}=\mathcal{S}(N,\mu,m),\quad v\in\{p,s\},
\label{kmmodel}
\end{align}
with a corresponding PDF defined as \cite{j:PlazaolaParis2017,j:MartinezParisJerez2017}
\begin{align}
f_{|h_{v}|^{2}}(x)=\sum^{N}_{j=0}\frac{C_{j}x^{m_{j}-1}\exp\left(-\frac{x}{\Omega_{j}}\right)}{\Omega^{m_{j}}_{j}(m_{j}-1)!},
\label{kmpdf}
\end{align}
while $\mu\leq m$ are specific integer parameters, which are explicitly defined in \cite[Table I]{j:MartinezParisJerez2017}, $N\triangleq m-\mu$, $m_{j}\triangleq m-j$, $\Omega_{j}\triangleq (\mu K+m)/(m \mu (1+K))$ with $K$ denoting the Rician$-K$ factor and $C_{j}\triangleq \binom{m-\mu}{j}(\frac{m}{\mu K+m})^{j}(\frac{\mu K}{\mu K+m})^{m-\mu-j}$. It is noteworthy that the $\mu$ parameter in \eqref{kmmodel} reflects the number of Tx antenna elements. In particular, in the case of a single-antenna PB (see Fig.~\ref{fig1}a), $\mu=1$ and $|h_{p}|^{2}\overset{\text{d}}=\mathcal{S}(N,1,m)$, while in the case of a $L-$antenna PB then $\mu=L$ and hence the effective received channel gain becomes $\|\mathbf{h}_{p}\|^{2}\overset{\text{d}}=\mathcal{S}(N,L,m)$. 

Some other key statistic results directly arising from \eqref{kmpdf} and are used in the rest of this paper are expressed as
\begin{align}
F_{|h_{v}|^{2}}(x)&=1-\sum^{N}_{j=0}\sum^{m_{j}-1}_{r=0}\frac{C_{j}\left(\frac{x}{\Omega_{j}}\right)^{r}}{r!}\exp\left(-\frac{x}{\Omega_{j}}\right),\label{cdfkm}\\
\mathbb{E}\left[|h_{v}|^{2}\right]&=1,\\
\mathbb{E}\left[{\rm ln}\left(|h_{v}|^{2}\right)\right]&=\sum^{N}_{j=0}C_{j}\left[\psi(m_{j})-{\rm ln}\left(\frac{1}{\Omega_{j}}\right)\right].
\label{kmexpectations}
\end{align}

\section{Performance Metrics}
According to the proposed mode-of-operation, it is naturally desirable to provide energy efficiency to ST, i.e., the EH phase in each time instance should be enough to transmit in every frame. Doing so, the total system throughput is also maximized since ST never stays inactive at a given frame. To this end, it is quite important to bound the effective range/distance between the $\rm PB-ST$ channel link in order to satisfy this condition.

\subsection{Effective EH Range} 
To derive the effective $d_{\rm PBST}$, according to \eqref{EH}, $\eta (1-\tau)P_{b}|h_{p}|^{2}[n]d^{-\alpha}_{\rm PBST}\geq \tau M$ should hold. Conditioned on the latter constraint, the average channel capacity reads as $\overline{C}\triangleq \mathbb{E}[\tau {\rm log}_{2}(1+\frac{\eta (1-\tau)P_{b}|h_{p}|^{2}|h_{s}|^{2}}{\tau N_{0} d^{\alpha}_{\rm PBST}d^{\alpha_{s}}_{\rm STSR}})]$. Further, it holds that $\overline{C}\geq C_{L}$, where $C_{L}$ denotes a lower bound of $\overline{C}$. Using the fact that ${\rm log}_{2}(1+\exp(z))$ is convex with respect to $z\quad \forall z>0$, \eqref{kmexpectations}, and based on Jensen's inequality, we get
\begin{align}
\nonumber
&C_{L}=\tau {\rm log}_{2}\Bigg(1+\frac{\eta (1-\tau)P_{b}\exp\left(\scriptstyle \mathbb{E}\left[{\ln}\left(|h_{p}|^{2}\right)\right]+\mathbb{E}\left[{\ln}\left(|h_{s}|^{2}\right)\right]\right)}{\tau N_{0} d^{\alpha}_{\rm PBST}d^{\alpha_{s}}_{\rm STSR}}\Bigg)\\
&=\tau {\rm log}_{2}\Bigg(1+\frac{\eta (1-\tau)P_{b}\exp\left(\scriptstyle 2\sum^{N}_{j=0}C_{j}\left[\psi(m_{j})-{\rm ln}\left(\frac{1}{\Omega_{j}}\right)\right]\right)}{\tau N_{0} d^{\alpha}_{\rm PBST}d^{\alpha_{s}}_{\rm STSR}}\Bigg).
\label{cl}
\end{align} 
On the other hand, the actual average capacity experienced at SR (using a fixed Tx power $M$ at ST) is given by $\overline{C}^{A}\triangleq \mathbb{E}[\tau {\rm log}_{2}(1+\frac{M |h_{s}|^{2}}{N_{0} d^{\alpha_{s}}_{\rm STSR}})]$, where it holds that $\overline{C}^{A}\geq C^{A}_{L}$ with $C^{A}_{L}$ denoting the lower bound of the benchmark average channel capacity $\overline{C}^{A}$. Following a similar methodology as previously, we get
\begin{align}
\overline{C}^{A}=\tau {\rm log}_{2}\Bigg(1+\frac{M\exp\left(\sum^{N}_{j=0}C_{j}\left[\psi(m_{j})-{\rm ln}\left(\frac{1}{\Omega_{j}}\right)\right]\right)}{N_{0} d^{\alpha_{s}}_{\rm STSR}}\Bigg).
\label{clbench}
\end{align} 
Thereby, setting $C_{L}=C^{A}_{L}$, we can directly interrelate the lower achievable channel capacity of the secondary system (given a fixed Tx power $M$), with the effective distance of the ${\rm PB-ST}$ channel link. Doing so, the effective range of $d_{\rm PB-ST}$ becomes $d_{\rm PB-ST}\leq d^{\star}$ with
\begin{align}
d^{\star}\triangleq\left[\frac{\eta (1-\tau)P_{b}}{\tau M}\exp\left(\sum^{N}_{j=0}C_{j}\left[\psi(m_{j})-{\rm ln}\left(\frac{1}{\Omega_{j}}\right)\right]\right)\right]^{\frac{1}{\alpha}}.
\label{effectrange}
\end{align} 

\subsection{Outage Probability and Average Throughput}
Outage probability at the $n^{\rm th}$ time instance is defined as the probability that ST transmits yet the received SNR at SR is below a prescribed outage threshold, $\gamma_{\rm th}$, or the probability that ST does not transmit and utilizes solely EH. It is expressed as
\begin{align}
P^{(n)}_{\rm out}(\gamma_{\rm th})\triangleq P_{\rm tr}[n]F_{\rm SNR}(\gamma_{\rm th})+(1-P_{\rm tr}[n]),
\label{poutdef}
\end{align} 
where $P_{\rm tr}[n]$ denotes the transmission probability at the $n^{\rm th}$ time instance, $\gamma_{\rm th}\triangleq 2^{\mathcal{R}}-1$ with $\mathcal{R}$ being the target (normalized) data rate for the secondary service in bps/Hz and 
\begin{align}
F_{\rm SNR}(\gamma_{\rm th})={\rm Pr}\left[\frac{M |h_{s}|^{2}}{d^{\alpha_{s}}_{\rm STSR}N_{0}}\leq \gamma_{\rm th}\right]=F_{|h_{s}|^{2}}\left(\frac{\gamma_{\rm th}N_{0}}{M d^{-\alpha_{s}}_{\rm STSR}}\right).
\label{CDFSNR}
\end{align} 

The remaining issue in order to derive the outage probability is $P_{\rm tr}[\cdot]$. The transmission probability at the $n^{\rm th}$ time instance is modeled as
\begin{align}
\nonumber
&P_{\rm tr}[n]\triangleq \Bigg\{{\rm Pr}\left[\eta P_{b}(1-\tau)\frac{|h_{p}|^{2}[n-1]}{d^{\alpha}_{\rm PBST}}+(1-\tau)M\geq M\right]\\
\nonumber
&\times {\rm Pr}\left[d_{\rm PBST}\leq d^{\star}\right]\Bigg\}+\Bigg\{{\rm Pr}\left[d_{\rm PBST}>d^{\star}\right]\\
&\times \Bigg({\rm Pr}\left[\eta P_{b}\frac{|h_{p}|^{2}[n-1]}{d^{\alpha}_{\rm PBST}}+(1-\tau)M\geq M\right]+\mathcal{J}(l)\Bigg)\Bigg\},
\label{ptr}
\end{align} 
where 
\begin{align}
\nonumber
\mathcal{J}(l)\triangleq {\rm Pr}\Bigg[&\left(\eta P_{b}\frac{\sum^{n-1}_{i=n-1-l}|h_{p}|^{2}[i]}{d^{\alpha}_{\rm PBST}}+(1-\tau)M\geq M\right)\&\\
&\left(\eta P_{b}\frac{\sum^{n-2}_{i=n-1-l}|h_{p}|^{2}[i]}{d^{\alpha}_{\rm PBST}}+(1-\tau)M<M\right)\Bigg],
\label{Jdef}
\end{align} 
and $l$ stands for the number of previous consecutive time instances where ST has been inactive (not transmitting) and used solely EH instead. The first term of \eqref{ptr} in curly brackets denotes the probability that ST is inside the effective $\rm PB-ST$ range. Therefore, the EH activity during the remaining fraction at the $(n-1)^{\rm th}$ time instance (along with the remaining energy $(1-\tau)M$ after transmission) is enough to meet the prescribed standards of the energy buffer and one-shot policy.

On the other hand, the second term of \eqref{ptr} in curly brackets denotes the case when the $\rm PB-ST$ link distance is outside the effective range. In this case, transmission in every consecutive time instance is not satisfied because the received harvested energy at ST does not meet the aforementioned condition due to severe channel fading and/or strong propagation attenuation. There are two possible scenarios: (\emph{i}) ST does not transmit at the previous $(n-1)^{\rm th}$ time instance, whereas it uses the entire duration for EH, and the received energy is enough (i.e., $\geq M$) to transmit at the $n^{\rm th}$ time instance; (\emph{ii}) The harvested energy was not enough for $l-1$ consecutive time instances, while it satisfies the aforementioned energy standards for $l$ ones, up to the $(n-1)^{\rm th}$ time instance. 

The latter scenario (\emph{ii}) is modeled by $\mathcal{J}(l)$, which includes two ordered and mutually correlated RVs, such that $\sum^{n-1}_{i=n-1-l}|h_{p}|^{2}[i]=\sum^{l}_{i=1}|h_{p}|^{2}[i]$ and $\sum^{n-2}_{i=n-1-l}|h_{p}|^{2}[i]=\sum^{l-1}_{i=1}|h_{p}|^{2}[i]$, while $\sum^{l-1}_{i=1}|h_{p}|^{2}[i]\leq \sum^{l}_{i=1}|h_{p}|^{2}[i]$.\footnote{It is assumed that small-scale channel fading coefficients are mutually independent across different time instances.} Hence, a similar approach as in \cite[Eq. (14) and Fig.~3a]{j:YangGSC} can be used to show that 
\begin{align}
\nonumber
&\mathcal{J}(l)=\\
&F_{\sum^{l-1}_{i=1}|h_{p}|^{2}[i]}\left(\frac{\tau M}{\eta P_{b}d^{-\alpha}_{\rm PBST}}\right)-F_{\sum^{l}_{i=1}|h_{p}|^{2}[i]}\left(\frac{\tau M}{\eta P_{b}d^{-\alpha}_{\rm PBST}}\right).
\label{Jdeff}
\end{align} 
Recall that $\sum^{l-1}_{i=1}|h_{p}|^{2}[i]\overset{\text{d}}=\mathcal{S}(N,l-1,m)$ and $\sum^{l}_{i=1}|h_{p}|^{2}[i]\overset{\text{d}}=\mathcal{S}(N,l,m)$. Thus, \eqref{Jdeff} can be easily computed via \eqref{cdfkm}, given a certain value of $l\geq 2$.

In addition, assuming that the random placement of ST within a given range follows the uniform (spatial) distribution, the PDF of $d_{\rm PBST}$ is given by
\begin{align}
f_{d_{\rm PBST}}(x)=\left\{
\begin{array}{ll}
\frac{2 x}{d^{2}_{\max}-d^{2}_{\min}}, &{\rm when }\quad d_{\min}\leq x\leq d_{\max},\\\\
0,& {\rm otherwise},
\end{array}
\right.
\label{distpdf}
\end{align}
where $d_{\min}$ denotes a minimum allowable distance from PB (mainly for human safety reasons due to the high radiated power of PB \cite[Remark~4]{j:XiaAissa2015}; otherwise, $d_{\min}=0$) and $d_{\max}$ is the radius of coverage (circular) area of PB.

Capitalizing on the above results, the first term in curly brackets of \eqref{ptr} can be rewritten as
\begin{align}
\left\{
\begin{array}{ll}
\int^{d^{\star}}_{d_{\min}}\overline{F}_{|h_{p}|^{2}}\left(\frac{\tau M x^{\alpha}}{\eta P_{b}(1-\tau)}\right)f_{d_{\rm PBST}}(x)dx,&\quad d_{\min}\leq d^{\star},\\\\
0,& \ \ \ {\rm otherwise}.
\end{array}
\right.
\label{term1}
\end{align}
Similarly, the second term in curly brackets reads as
\begin{align}
&\left\{
\begin{array}{ll}
\int^{d_{\max}}_{d^{\star}}\left[\scriptstyle \overline{F}_{|h_{p}|^{2}}\left(\frac{\tau M x^{\alpha}}{\eta P_{b}}\right)+\mathcal{J}(l)\right]f_{d_{\rm PBST}}(x)dx,&d^{\star}\leq d_{\max},\\\\
0,& {\rm otherwise}
\end{array}
\right.\label{term22}\\
&\approx \left\{
\begin{array}{ll}
\int^{d_{\max}}_{d^{\star}}\scriptstyle \overline{F}_{|h_{p}|^{2}}\left(\frac{\tau M x^{\alpha}}{\eta P_{b}}\right)f_{d_{\rm PBST}}(x)dx,&d^{\star}\leq d_{\max},\\\\
0,& {\rm otherwise},
\end{array}
\right.
\label{term2}
\end{align}
where \eqref{term2} follows by neglecting $\mathcal{J}(l)$ from \eqref{term22}. For a numerical setting of practical interest (provided in the next section), $\mathcal{J}(l)$ takes values in the order of $10^{-6}$ by setting $l=2$, while it takes values $\ll 10^{-8}$ for $l=3$. On the other hand, evaluating \eqref{term2} yields values $\gg 10^{-3}$. Hence, the approximation of \eqref{term2} is very tight.

Using \eqref{cdfkm} and \eqref{distpdf}, \eqref{term1} and \eqref{term2} are, respectively, evaluated with the aid of \cite[Eq. (3.381.9)]{tables} as
\begin{align}
\nonumber
&\Phi_{1}\triangleq\\
&\left\{
\begin{array}{l}
\sum^{N}_{j=0}\sum^{m_{j}-1}_{r=0}\frac{2 C_{j}\left(\frac{\tau M}{\eta P_{b}(1-\tau)\Omega_{j}}\right)^{-\frac{2}{\alpha}}}{r!\alpha \left((d^{\star})^{2}-d^{2}_{\min}\right)}\\
\times \left[\Gamma\left(r+\frac{2}{\alpha},\frac{\tau M d^{\alpha}_{\min}}{\eta P_{b}(1-\tau)\Omega_{j}}\right)-\Gamma\left(r+\frac{2}{\alpha},\frac{\tau M (d^{\star})^{\alpha}}{\eta P_{b}(1-\tau)\Omega_{j}}\right)\right],\\\\
\quad\quad\quad\quad\quad\quad\quad\quad\quad\quad\quad\quad\quad\quad\quad\quad\quad\quad d_{\min}\leq d^{\star},\\\\
0,\quad {\rm otherwise},
\end{array}
\right.
\label{term1cl}
\end{align}
and
\begin{align}
\nonumber
&\Phi_{2}\triangleq\\
&\left\{
\begin{array}{l}
\sum^{N}_{j=0}\sum^{m_{j}-1}_{r=0}\frac{2 C_{j}\left(\frac{\tau M}{\eta P_{b}\Omega_{j}}\right)^{-\frac{2}{\alpha}}}{r!\alpha \left(d^{2}_{\max}-(d^{\star})^{2}\right)}\\
\times \left[\Gamma\left(r+\frac{2}{\alpha},\frac{\tau M (d^{\star})^{\alpha}}{\eta P_{b}\Omega_{j}}\right)-\Gamma\left(r+\frac{2}{\alpha},\frac{\tau M d^{\alpha}_{\max}}{\eta P_{b}\Omega_{j}}\right)\right],\\\\
\quad\quad\quad\quad\quad\quad\quad\quad\quad\quad\quad\quad\quad\quad\quad\quad\quad\quad d^{\star}\leq d_{\max},\\\\
0,\quad {\rm otherwise}.
\end{array}
\right.
\label{term2cl}
\end{align}
Finally, by inserting \eqref{term1cl} and \eqref{term2cl} into \eqref{ptr}, we get $P_{\rm tr}[n]\approx \Phi_{1}+\Phi_{2}$. Also, inserting \eqref{CDFSNR} into \eqref{poutdef}, the outage probability is derived in a closed formula.

Furthermore, the average effective throughput (in bps/Hz) is defined as
\begin{align}
\overline{T}\triangleq \tau \mathcal{R}\left[1-P_{\rm out}\left(2^{\mathcal{R}}-1\right)\right],
\label{avt}
\end{align}
which can be directly computed via \eqref{poutdef}.

\subsection{Practical Energy Harvesting Imperfections}
In practice, the EH and power amplifier circuits may not perform ideally. In this case, for a given Tx power at ST, i.e., $M$, the corresponding power amplifier consumes a total power $\rho M$, where $\rho\geq 1$ denotes the power amplifier inefficiency factor. In addition, the EH circuit consumes a constant power $P_{c}$ during transmission \cite{j:MorsiMichalopoulosSchober2018}. Hence, according to the proposed mode-of-operation, ST transmits using a constant power $M$ whenever its stored energy reaches $\overline{M}\triangleq \rho M+P_{c}$; otherwise it does not transmit and performs EH instead. Thus, in the case of EH imperfections, the transmission probability in \eqref{ptr} becomes 
\begin{align}
\nonumber
&P_{\rm tr}[n]\approx \Bigg\{{\rm Pr}\left[\eta P_{b}(1-\tau)\frac{|h_{p}|^{2}[n-1]}{d^{\alpha}_{\rm PBST}}+(1-\tau)\overline{M}\geq \overline{M}\right]\\
\nonumber
&\times {\rm Pr}\left[d_{\rm PBST}\leq d^{\star}\right]\Bigg\}+\Bigg\{{\rm Pr}\left[d_{\rm PBST}>d^{\star}\right]\\
&\times {\rm Pr}\left[\eta P_{b}\frac{|h_{p}|^{2}[n-1]}{d^{\alpha}_{\rm PBST}}+(1-\tau)\overline{M}\geq \overline{M}\right]\Bigg\}.
\label{ptrimp}
\end{align} 
To this end, after some simple manipulations, we can easily show that the previously presented analytical results for the considered performance metrics are valid in this case, by substituting $M$ with $\overline{M}$ in \eqref{term1cl} and \eqref{term2cl}.  

\section{Numerical Results and Concluding Remarks}
In this section, numerical results are presented and compared with Monte-Carlo simulation results, which are illustrated as line-curves and circle-marks, respectively. In the ensuing simulation experiments, the Tx power at PB is set to be $P_{b}=33$dBm, the Tx power at ST is $M=20$dBm, $\{\alpha = 2.4, m=20, K=7\}$, which reflect line-of-sight propagation conditions for the $\rm PB-ST$ channel link \cite{j:MartinezParisJerez2017,j:PlazaolaParis2017}, the power-to-energy conversion efficiency is $\eta=0.85$ \cite{j:XiaAissa2015}, and $N_{0}=-101$dBm. For the ${\rm ST-SR}$ signal path loss, $a_{s}=3$ is assumed. Also, for the scenario of EH imperfection, $\rho=1.2$ and $P_{c}=-30$dBm (i.e., $\overline{M}\approx 0.12$mJ for $T=1$ms). Finally, $\{$$d_{\min}=1$m and $d_{\max}=15$m$\}$, which correspond to the coverage of the $\rm PB-ST$ channel link, while $d_{\rm STSR}=30$m. 

\begin{figure}[!t]
\centering
\includegraphics[trim=1.8cm 0.0cm 2.5cm 0cm, clip=true,totalheight=0.18\textheight]{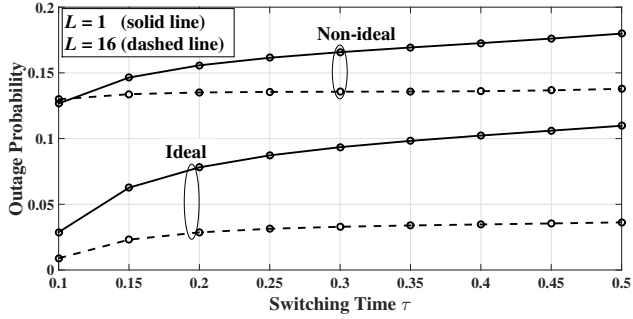}
\caption{Outage probability as per \eqref{poutdef} vs. various values of the switching time $\tau$ for a single-antenna PB (i.e. $L=1$) and for a multi-antenna PB (i.e., $L=16$), when $\mathcal{R}=1$bps/Hz.}
\label{fig2}
\end{figure}

In Fig.~\ref{fig2}, the outage performance of the considered secondary system is presented for two different setups; the ideal EH case (where $\rho=1$ and $P_{c}=0$) and the non-ideal case reflecting the practical imperfection of EH circuit. Obviously, the outage performance is worse for a higher switching time $\tau$. This occurs since ST transmits a longer duration in each time instance for higher $\tau$, which corresponds to more energy losses of the energy buffer at ST. In turn, the probability that ST stays inactive (i.e., non-transmitting) in a given time instance is increased, which also results to an increased probability of an outage event. It is also noteworthy that the case when a multi-antenna PB is adopted drastically impacts on the enhancement of the outage performance since the amount of collected energy at ST is proportionally enhanced; hence, the probability that ST transmits using a constant Tx power $M$, given a full energy buffer, is correspondingly increased.  

\begin{figure}[!t]
\centering
\includegraphics[trim=1.8cm 0.0cm 2.5cm 0cm, clip=true,totalheight=0.18\textheight]{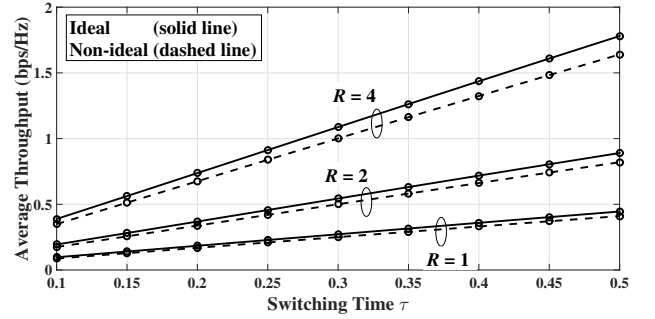}
\caption{Average effective throughput as per \eqref{avt} vs. various values of the switching time $\tau$ for a single-antenna PB and for three different scenarios of target data rate $\mathcal{R}$ (in bps/Hz).}
\label{fig3}
\end{figure}

In Fig.~\ref{fig3}, the average effective throughput of the considered system is presented, which is enhanced for higher $\tau$ values. As expected, this is a reasonable outcome since for a longer transmit duration, in each time instance, more data streams can potentially be transferred from ST to SR. As noticed from Fig.~\ref{fig2} and also verified in Fig.~\ref{fig3}, the efficiency of EH circuit (i.e., the quality of its hardware gear) plays a key role to the overall performance of the considered system. To this end, network practitioners working into this field should pay attention to the efficient design of EH equipment to maintain sufficient communication quality in EH-enabled transmission systems.

\bibliographystyle{IEEEtran}
\bibliography{IEEEabrv,References}

\begin{thebibliography}{10}
\providecommand{\url}[1]{#1}
\csname url@samestyle\endcsname
\providecommand{\newblock}{\relax}
\providecommand{\bibinfo}[2]{#2}
\providecommand{\BIBentrySTDinterwordspacing}{\spaceskip=0pt\relax}
\providecommand{\BIBentryALTinterwordstretchfactor}{4}
\providecommand{\BIBentryALTinterwordspacing}{\spaceskip=\fontdimen2\font plus
\BIBentryALTinterwordstretchfactor\fontdimen3\font minus
  \fontdimen4\font\relax}
\providecommand{\BIBforeignlanguage}[2]{{%
\expandafter\ifx\csname l@#1\endcsname\relax
\typeout{** WARNING: IEEEtran.bst: No hyphenation pattern has been}%
\typeout{** loaded for the language `#1'. Using the pattern for}%
\typeout{** the default language instead.}%
\else
\language=\csname l@#1\endcsname
\fi
#2}}
\providecommand{\BIBdecl}{\relax}
\BIBdecl

\bibitem{j:KansalSadaf2007}
A.~Kansal, J.~Hsu, S.~Zahedi, and M.~B. Srivastava, ``Power management in
  energy harvesting sensor networks,'' \emph{ACM Trans. Embed. Comput. Syst.},
  vol.~6, no.~4, Sep. 2007.

\bibitem{j:FalkensteinCostinett2011}
E.~Falkenstein, D.~Costinett, R.~Zane, and Z.~Popovic, ``Far-field {RF}-powered
  variable duty cycle wireless sensor platform,'' \emph{IEEE Trans. Circuits
  and Systems II: Express Briefs}, vol.~58, no.~12, pp. 822--826, Dec. 2011.

\bibitem{j:XiaAissa2015}
M.~Xia and S.~A\"{i}ssa, ``On the efficiency of far-field wireless power
  transfer,'' \emph{{IEEE} Trans. Signal Process.}, vol.~63, no.~11, pp.
  2835--2847, Jun. 2015.

\bibitem{j:Haykin}
S.~Haykin, ``Cognitive radio: brain-empowered wireless communications,''
  \emph{{IEEE} J. Sel. Areas Commun.}, vol.~23, no.~2, pp. 201--220, Feb. 2005.

\bibitem{j:BockelmannPratas2016}
C.~Bockelmann, N.~Pratas, H.~Nikopour, K.~Au, T.~Svensson, C.~Stefanovic,
  P.~Popovski, and A.~Dekorsy, ``Massive machine-type communications in 5{G}:
  {P}hysical and mac-layer solutions,'' \emph{{IEEE} Commun. Mag.}, vol.~54,
  no.~9, pp. 59--65, Sep. 2016.

\bibitem{j:HuangHan2015}
X.~Huang, T.~Han, and N.~Ansari, ``On green-energy-powered cognitive radio
  networks,'' \emph{{IEEE} Commun. Surveys Tuts.}, vol.~17, no.~2, pp.
  827--842, Second Quarter 2015.

\bibitem{j:MiridakisTsiftsisAlexandropoulosDebbah2016}
N.~I. Miridakis, T.~A. Tsiftsis, G.~C. Alexandropoulos, and M.~Debbah, ``Green
  cognitive relaying: {O}pportunistically switching between data transmission
  and energy harvesting,'' \emph{{IEEE} J. Sel. Areas Commun.}, vol.~34,
  no.~12, pp. 3725--3738, Dec. 2016.

\bibitem{j:KuLiChen2016}
M.~L. Ku, W.~Li, Y.~Chen, and K.~J.~R. Liu, ``Advances in energy harvesting
  communications: {P}ast, present, and future challenges,'' \emph{{IEEE}
  Commun. Surveys Tuts.}, vol.~18, no.~2, pp. 1384--1412, Second Quarter 2016.

\bibitem{j:LeeZhang2017}
S.~Lee and R.~Zhang, ``Distributed wireless power transfer with energy
  feedback,'' \emph{{IEEE} Trans. Signal Process.}, vol.~65, no.~7, pp.
  1685--1699, Apr. 2017.

\bibitem{j:MorsiMichalopoulosSchober2018}
R.~Morsi, D.~Michalopoulos, and R.~Schober, ``Performance analysis of
  near-optimal energy buffer aided wireless powered communication,''
  \emph{{IEEE} Trans. Wireless Commun.}, To appear, 2018.

\bibitem{j:MartinezParisJerez2017}
F.~J. Lopez-Martinez, J.~F. Paris, and J.~M. Romero-Jerez, ``The $\kappa$-$\mu$
  shadowed fading model with integer fading parameters,'' \emph{{IEEE} Trans.
  Veh. Technol.}, vol.~66, no.~9, pp. 7653--7662, Sep. 2017.

\bibitem{j:PlazaolaParis2017}
\BIBentryALTinterwordspacing
U.~Fern{\'{a}}ndez{-}Plazaola, L.~Moreno{-}Pozas, F.~J.
  L{\'{o}}pez{-}Mart{\'{\i}}nez, J.~F. Paris, E.~Martos{-}Naya, and J.~M.
  Romero{-}Jerez, ``A tractable product channel model for line-of-sight
  scenarios,'' \emph{Under review}, 2017. [Online]. Available:
  \url{http://arxiv.org/abs/1711.05650}
\BIBentrySTDinterwordspacing

\bibitem{tables}
I.~S. Gradshteyn and I.~M. Ryzhik, \emph{Table of Integrals, Series, and
  Products}.\hskip 1em plus 0.5em minus 0.4em\relax Academic Press, 2007.

\bibitem{j:ZhangChen2017}
C.~Zhang and Y.~Chen, ``Wireless power transfer strategies for cooperative
  relay system to maximize information throughput,'' \emph{{IEEE} Access},
  vol.~5, pp. 2573--2582, 2017.

\bibitem{j:ZhaoLinKaragiannidis2018}
F.~Zhao, H.~Lin, C.~Zhong, Z.~Hadzi-Velkov, G.~K. Karagiannidis, and Z.~Zhang,
  ``On the capacity of wireless powered communication systems over {R}ician
  fading channels,'' \emph{{IEEE} Trans. Commun.}, vol.~66, no.~1, pp.
  404--417, Jan. 2018.

\bibitem{j:KestutisKeskinen2016}
K.~Grigoras, J.~Keskinen, L.~Grönberg, E.~Yli-Rantala, S.~Laakso,
  H.~Välimäki, P.~Kauranen, J.~Ahopelto, and M.~Prunnila, ``Conformal
  titanium nitride in a porous silicon matrix: {A} nanomaterial for in-chip
  supercapacitors,'' \emph{Nano Energy}, vol.~26, pp. 340 -- 345, 2016.

\bibitem{j:YangGSC}
H.-C. Yang, ``New results on ordered statistics and analysis of
  minimum-selection generalized selection combining ({GSC}),'' \emph{{IEEE}
  Trans. Wireless Commun.}, vol.~5, no.~7, pp. 1876--1885, Jul. 2006.

\end{thebibliography}

\vfill

\end{document}